\documentclass[prd,floatfix, nofootinbib,showpacs,
preprintnumbers
]{revtex4}
\usepackage{amsmath}
\usepackage{amssymb}
\usepackage{epsfig}
\usepackage{graphicx}
\usepackage{color}
\usepackage{hyperref}
\usepackage{dsfont}
\usepackage{wasysym}
\newcommand{\be}{\begin{equation}}
\newcommand{\ee}{\end{equation}}
\newcommand{\bdm}{\begin{displaymath}}
\newcommand{\edm}{\end{displaymath}}
\newcommand{\bea}{\begin{eqnarray}}
\newcommand{\eea}{\end{eqnarray}}
\newcommand{\nn}{\nonumber}

\newcommand{\vev}[1]{\langle #1 \rangle}

\newcommand{\PMNS}{V_{PMNS}}
\newcommand{\PMNSx}{V}


\begin{document}

\title{Witten's loop in the flipped $SU(5)$ unification}


\author{Michal Malinsk\'{y}\footnote{Presenting author; email:malinsky@ipnp.troja.mff.cuni.cz}\hskip 2mm}
\affiliation{Institute of Particle and Nuclear Physics,
Faculty of Mathematics and Physics,
Charles University in Prague, V Hole\v{s}ovi\v{c}k\'ach 2,
180 00 Praha 8, Czech Republic
}

\author{Carolina Arbel\'aez Rodr\'iguez}\affiliation{Centro de F\'isica Te\'orica de Part\'iculas, Instituto Superior T\'ecnico, Universidade T\'ecnica de Lisboa, Av. Rovisco Pais 1, 1049-001, Lisboa Portugal \\ and \\AHEP Group, Instituto de F\'\i sica Corpuscular -- C.S.I.C./Universitat de Val\`encia Edificio de Institutos de Paterna, Apartado 22085, E--46071 Val\`encia, Spain
}

\author{Helena Kole\v{s}ov\'a}\affiliation{Institute of Particle and Nuclear Physics,
Faculty of Mathematics and Physics,
Charles University in Prague, V Hole\v{s}ovi\v{c}k\'ach 2,
180 00 Praha 8, Czech Republic \\ and \\Faculty of Nuclear Sciences and Physical Engineering, Czech Technical University in Prague, B\v{r}ehov\'a 7, \\ 115 19 Praha 1, Czech Republic
}

\begin{abstract}
We study a very simple, yet potentially realistic renormalizable flipped SU(5) scenario in which the right-handed neutrino masses are generated at very high energies by means of a two-loop diagram similar to that identified by E. Witten in the early 1980's in the SO(10) GUT framework. This mechanism leaves its traces in the baryon number violating signals such as the proton decay, especially in the ``clean'' channels with a charged lepton and a neutral meson in the final state.  
\end{abstract}

\maketitle

\section{Introduction}
Besides the canonical implementation of the seesaw mechanism \cite{Minkowski:1977sc,Yanagida:1979as,Mohapatra:1979ia,Schechter:1980gr,Lazarides:1980nt,Foot:1988aq} exploiting the three inequivalent tree-level renormalizable openings of the Weinberg's operator $LLHH$ at a certain very high scale, the unprecedented smallness of the light neutrino masses indicated by the beta-decay and cosmology data is often attributed to (multi-) loop suppression of Feynman diagrams featuring new physics at relatively low energies (see, e.g., \cite{Zee:1980ai,Zee:1985rj,Zee:1985id,Babu:1988ki}), often in the TeV ballpark. Recently, a lot of studies focusing on distinctive features of various such low-scale models (cf. \cite{Bonnet:2012kz,Angel:2012ug,Babu:2013pma} and references therein) has appeared and their testability at the LHC and other facilities~\cite{Baek:2012ub,Ohlsson:2009vk,Nebot:2007bc,AristizabalSierra:2006gb,Frampton:2001eu} has been discussed thoroughly. 

Should proton decay be found in the next generation of megaton-scale facilities such as Hyper-Kaminokande or LBNE~\cite{Abe:2011ts,Akiri:2011dv,Autiero:2007zj} a qualitatively new window on this conundrum will wide open; this concerns namely the potential testability of those models in which the perturbative lepton number violation behind the Weinberg's operator is tied to the violation of baryon number in a simple way, typically, by means of new interactions inherent to some kind of a unified theory. 

The Witten's loop mechanism~\cite{Witten:1979nr} for the radiative generation of the right-handed (RH) Majorana neutrino mass ($M_{\nu}^{M}$) in the simplest $SO(10)$ grand unified theories (GUTs) is a standard example of such a twist; the relevant two-loop Feynman diagrams make use of the baryon and lepton number violating gauge and scalar interactions giving mass to the RH neutrinos in a framework where the relevant tree-level contraction (including a Lorentz scalar that transforms as a 126-dimensional $SO(10)$ tensor) is unavailable. 

Unfortunately, soon after its invention the Witten's mechanism has been mostly abandoned as a mere curiosity. Among the main reasons there was namely the tension between the gauge unification constraints which, in the non-supersymmetric theories, require the rank-breaking vacuum expectation value (VEV) to be well within the GUT ``desert''~\cite{Chang:1984qr,Deshpande:1992au,Deshpande:1992em,Bertolini:2009qj} (which, however, leads to an ``oversuppression'' of thus calculated $M_{\nu}^{M}$), and the general tendency of supersymmetric theories to cancel the GUT-scale $F$-type loop diagrams (with exceptions like, for instance, the works~\cite{Bajc:2004hr,Bajc:2005aq} in the split-SUSY context where such a cancellation has been tamed by pushing the masses of the scalar superpartners up to the very GUT scale).  

A possible way out that we would like to entertain in this study consists in a ``controlled'' departure from the strict gauge unification constraints inherent to grand unifications with a clear objective to push the rank- (i.e., the lepton-number-) breaking VEV as high as possible, i.e., to the typical GUT-scale ballpark.  In particular, we shall try to exploit the variant(s) of the Witten's loop in gauge unifications that are not ``grand'', i.e.,  those that are not based on a simple  gauge group. At the same time, we shall be interested only in those scenarios whose gauge group can be embedded into the original Witten's $SO(10)$ and in which the perturbative BNV signals could be of any relevance, i.e., those models that may be constrained from such kind of physics. 

Remarkably enough, there is a single renormalizable gauge extension of the Standard model that obeys all these requirements, namely, the flipped $SU(5)$ scenario~\cite{Derendinger:1983aj,PhysRevLett.45.413,Barr:1981qv}, cf. also~\cite{Rodriguez:2013rma}. In this framework, the quarks and leptons of the Standard Model (SM) plus the mandatory RH neutrino $\nu^{c}$ are embedded into three irreducible representations of the $SU(5)\otimes U(1)_{X}$ subgroup of $SO(10)$, namely\footnote{Note that, up to an overall normalization, the $X$-charges of this set of fields are fixed by the condition of gauge anomaly cancellation.}
$\overline{5}_{M}\equiv (\overline{5},-3)$ containing $u^{c}$ and $L$, $10_{M}\equiv (10,+1)$ accommodating $d^{c}$, $Q$ and $\nu^{c}$ and $1_{M}\equiv (1,+5)$ corresponding to $e^{c}$ (all fields left-handed). Note that with such an assignment the SM hypercharge can be spanned over both gauge factors as $Y=\tfrac{1}{5}(X-T_{24})$ where $T_{24}$ corresponds to the ``standard'' $SU(5)$ hypercharge in the SM normalization. Hence, the SM effective coupling $g'$ is matched to a linear combination of the ``unified''  coupling $g_{5}$
associated to $SU(5)$ and the a-priori unknown $g_{X}$ coupling of $U(1)_{X}$ and, as such, it may yield the correct low-scale value even in the ``desert'' picture without low-energy supersymmetry. The  $SU(5)\otimes U(1)_{X}$ gauge symmetry is broken down to the SM by means of a VEV of a 10-dimensional scalar $10_{H}$ transforming as $ (10,+1)$ while the electroweak symmetry breaking is provided by the traditional Higgs doublet contained in  ${5}_{H}\equiv ({5},-2)$.
The renormalizable Yukawa Lagrangian
\be\label{welcome2}
{\cal L}_{Y}\ni Y_{10}10_{M}10_{M}5_{H}+Y_{\overline{5}}10_{M}\overline{5}_{M}5^{*}_{H}+Y_{1}\overline{5}_{M}1_{M}5_{H}+h.c.
\ee
then provides $M_{\nu}^{D}=M_{u}^{T}$, $M_{d}=M_{d}^{T}$ and an arbitrary $M_{e}$ which is very welcome as none of these correlations conflicts with the  observed quark and lepton flavour pattern (as does $M_{d}=M_{e}^{T}$ in the ``standard'' $SU(5)$).

Furthermore, there are several distinctive features in the BNV signals in the flipped $SU(5)$ that are relatively easy to be distinguished from those typical to other unified scenarios. Besides the generally high predictivity for the $d=6$ proton decay into antineutrinos (shared with some other scenarios)\footnote{This owes namely to the fact that there is typically a single effective operator governing these channels and the option to get rid of the uncertainties in the relevant flavour rotations by summing over the final neutrino states.} which is, in the minimal case,  demonstrated by a firm prediction $\Gamma(p\to K^+\overline{\nu})=0$, the flipped $SU(5)$ offers a relatively good grip on the $p$-decay with a neutral meson and a charged lepton in the final state that are typically much easier to look for in the megaton-scale water-Cherenkow~\cite{Abe:2011ts}/liquid Argon~\cite{Akiri:2011dv}/liquid scintilator~\cite{Autiero:2007zj} environment.    
In particular, one can write
\be\label{gamma3}
\frac{\Gamma(p\to \pi^0 e_\alpha^+)}{\Gamma(p\to \pi^+\overline{\nu})}=\frac{1}{2}|(V_{CKM})_{11}|^2|(U_e^L)_{\alpha 1}|^2\,,
\quad\frac{\Gamma(p\to \eta e_\alpha^+)}{\Gamma(p\to \pi^+\overline{\nu})}=\frac{C_2}{C_1}|(V_{CKM})_{11}|^2|(U_e^L)_{\alpha 1}|^2\,,\ee
\be\label{gamma5}
\frac{\Gamma(p\to K^0 e_\alpha^+)}{\Gamma(p\to \pi^+\overline{\nu})}=\frac{C_3}{C_1}|(V_{CKM})_{12}|^2|(U_e^L)_{\alpha 1}|^2\,,
\ee
where $U_e^L$ is the LHS diagonalization matrix in the charged lepton sector, $V_{CKM}$ is the Cabibbo-Kobayashi-Maskawa matrix and $
C_1=m_p A_L^2|\alpha|^2(1+D+F)^2/8\pi f_\pi^2$, $
C_2=(m_p^2-m_\eta^2)^2 A_L^2|\alpha|^2(1+D-3F)^2/48\pi m_p^3 f_\pi^2$ and $
C_3=(m_p^2-m_K^2)^2A_L^2|\alpha|^2\left[1+\frac{m_p}{m_B}(D-F)\right]^2/8\pi m_p^3 f_\pi^2$
are long-distance factors; for more detail see~\cite{Nath:2006ut}.
 The denominator in formulae (\ref{gamma3})-(\ref{gamma5}) is given by
$
\Gamma(p\to \pi^+\overline{\nu})=C_1 \left(\frac{g_G}{M_{G}}\right)^4
$ where $M_{G}$ stands for the mass of the heavy gauge bosons and $g_{G}$ is the unified non-abelian gauge coupling.

The whole point is that the $U_e^L$ matrix in Eqs.~(\ref{gamma3})-(\ref{gamma5}) may be written as $\PMNS U_\nu$ where $\PMNS$ is the lepton flavour mixing matrix measurable in neutrino experiments and $U_\nu$ is the diagonalization matrix in the sector of the light neutrinos that one may get a grip on from the Witten's mechanism. Let us note that this is impossible in the ``usual'' approach to the renormalizable flipped $SU(5)$ (i.e., in the models where the RH neutrino masses are generated via  
an extra scalar representation transforming as a 50-dimensional four-index tensor $50_{H}=(50,-2)$ coupled to the fermionic $10_{M}\otimes 10_{M}$ bilinear, see, e.g.,~\cite{Das:2005eb}, or via extra matter singlets, cf.~\cite{Abel:1989hq} and references therein); there the $U_\nu$ matrix remains essentially unconstrained.
\section{Witten's loop in the flipped $SU(5)$ unification}
\begin{figure}[t]
\parbox{6cm}{\includegraphics[width=6cm,height=3.7cm]{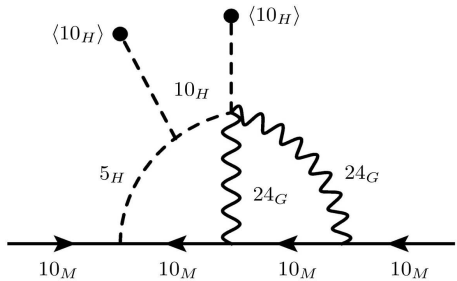}}
\caption{The gauge structure of the Witten's loop in the flipped $SU(5)$ scenario under consideration. Note that we display just one representative out of several graphs that may be obtained from the one above by permutation of the internal lines.}
\label{figwitten}
\end{figure}

In the model under consideration the RH neutrino masses are generated at two loops by a diagram depicted in FIG.~\ref{figwitten}. Obviously, the pair of the adjoint gauge fields $24_{G}$ together with the $5_{H}$ scalar are arranged in just the right way to mimic the desired insertion of the VEV of an effective $(50,-2)$ at the renormalizable level.    
These graphs can be evaluated readily: 
\be
M_{\nu}^{M}= \left(\frac{1}{16\pi^{2}}\right)^{2}g_{G}^{4}Y_{10}\,\mu\frac{\vev{10_{H}}^{2}}{M_{G}^{2}}\times {\cal O}(1)\label{MnuMmain}\,,
\ee
where $\mu$ is the (dimensionful) trilinear scalar coupling among $10_{H}$'s and $5_{H}$, $Y_{10}$ is the Yukawa coupling of $5_{H}$ to the matter bilinear $10_{M}\otimes 10_{M}$, $\vev{10_{H}}$ is the GUT-symmetry-breaking VEV and the ${\cal O}(1)$ factor stands for the remainder of the relevant expression which, besides the double loop-momentum integration  contains, for example, the unitary transformations among the defining and the mass bases. Since, however, the loop can not be evaluated without a detailed information about the heavy spectrum of the theory we shall just formally cancel $g_{G}^{2}V_{G}^{2}$ against the $M_{G}^{2}$ factor (assuming, as usual, $M_{G}= g_{G} V_{G}$ up to an order 1 constant) and rewrite Eq.~(\ref{MnuMmain}) as
\be
M_{\nu}^{M}= \left(\frac{1}{16\pi^{2}}\right)^{2}g_{G}^{2}Y_{10}\,\mu\, K\label{MnuMmain2}\,,
\ee
where the inaccuracy in the last step has been concealed\footnote{This, in fact, is the best one can do until all the scalar potential couplings are fixed.} into a (hitherto unknown) factor $K$. Assuming no accidental cancellations, a qualified guess~\cite{Babuprivateconversation} puts this factor to the ${\cal O}(10)$ domain; hence, in what follows we shall consider $K$ in the interval from $5$ to $50$ and cast all our results as functions of this parameter.
\section{The minimal model}
The remaining two parameters in Eq.~(\ref{MnuMmain2}), i.e., $Y_{10}$ and $\mu$, can be on very general grounds constrained from the requirements of the SM vacuum stability and perturbativity of the entire framework which we shall now elaborate on.
\subsection{General prerequisites}
\subsubsection{Vacuum stability constraints}
\noindent
With the scalar potential parametrized like 
\bea\label{potential}
V_0&=&\frac{1}{2}m_{10}^{2}{\rm Tr}(10_{H}^{\dagger}10_{H})+m_{5}^{2}5_{H}^{\dagger}5_{H}+\frac{1}{8}(\mu \varepsilon_{ijklm}10_{H}^{ij}10_{H}^{kl}5_{H}^{m}+h.c.)+\nn\\
&+&
\frac{1}{4}\lambda_{1}[{\rm Tr}(10_{H}^{\dagger}10_{H})]^{2}+
\frac{1}{4}\lambda_{2}{\rm Tr}(10_{H}^{\dagger}10_{H}10_{H}^{\dagger}10_{H})+
\lambda_{3}(5_{H}^{\dagger}5_{H})^{2}+
\frac{1}{2}\lambda_{4}{\rm Tr}(10_{H}^{\dagger}10_{H})(5_{H}^{\dagger}5_{H})+\lambda_{5}5_{H}^{\dagger}10_{H}10_{H}^{\dagger}5_{H}\nn\,,
\eea
the scalar spectrum of the theory may be calculated readily: 
\begin{align}
&m_{G_{1,\ldots,16}}^{2}=0\label{Goldstones}\\
&m_H^{2}=\left[4\lambda_{3}-\frac{2(\lambda_{4}+\lambda_{5})^{2}}{2\lambda_{1}+\lambda_{2}}\right]v^{2}\,, \label{Higgsmass}\\
&m_S^{2}=2(2\lambda_{1}+\lambda_{2})V_{G}^{2}\,,\nn\\
&m_{\Delta_1}^{2}=-\tfrac{1}{2}(\lambda_{2}+\lambda_{5})V_{G}^{2}-\tfrac{1}{2}V_{G}\sqrt{(\lambda_{2}-\lambda_{5})^{2}V_{G}^{2}+4\mu^{2}},\nn\\
&m_{\Delta_2}^{2}=-\tfrac{1}{2}(\lambda_{2}+\lambda_{5})V_{G}^{2}+\tfrac{1}{2}V_{G}\sqrt{(\lambda_{2}-\lambda_{5})^{2}V_{G}^{2}+4\mu^{2}}\nn.
\end{align}
Here $V_{G}$ stands for the high-scale VEV of $10_{H}$ and $v$ for the electroweak one. Note that the 16 zeroes correspond to the Goldstone bosons associated to the $SU(5)\otimes U(1)/SU(3)_{QCD}\otimes U(1)_{QED}$ coset, $H$ is the SM Higgs boson, $S$ is the heavy singlet survivor and $\Delta_{1,2}$ are the two different coloured triplet scalars. It is easy to see that $m^{2}_{\Delta_{1}}$ is tachyonic 
unless
\be
|\mu|< \sqrt{ \lambda_{2}\lambda_{5}}V_{G}\label{allimportant}\,,
\ee
which, in turn, may be viewed as the only domain of the parameter space that can, at the tree level, support a (locally) stable SM vacuum. Besides this, one should also have $2\lambda_{1}+\lambda_{2} >0$, $2\lambda_{3}(2\lambda_{1}+\lambda_{2}) > (\lambda_{4}+\lambda_{5})^{2}$ and $\lambda_{2}+\lambda_{5} < 0$.
\subsubsection{Perturbativity constraints}
\noindent For the sake of this study we shall implement a simplified set of perturbativity constraints in the form
\be\label{perturbativityconstraints}
|\lambda_{i}|\leq 4\pi \;\forall i\;\;,\;|(Y_{X})_{ij}|\leq 4\pi\;\; \text{for }X=10,\overline{5},1\text{ and }\forall i,j\,,
\ee
which are understood to be imposed on the running couplings at the unification scale assuming that their subsequent evolution to the electroweak scale is not pathological; for more detail see, e.g.,  the discussion in~\cite{Rodriguez:2013rma}. Besides that, one may also fix $4\lambda_{3}(2\lambda_{1}+\lambda_{2})-2(\lambda_{4}+\lambda_{5})^{2}=0$ which reflects the observed decline of the effective Higgs quartic coupling towards very high energies, see, e.g.,~\cite{Buttazzo:2013uya} and references therein. On the other hand, this extra constraint does not add much to the discussion below so we shall not further elaborate on it.
\subsubsection{The central formula}
\noindent With all this at hand, the Witten's formula (\ref{MnuMmain2}) may be finally recast in the form 
\be\label{centralformula}
\max_{i,j\in\{1,2,3\}}|(D_u U_\nu^\dagger D_\nu^{-1} U_\nu^* D_u)_{ij}|\leq \frac{\alpha_{G}}{4\pi}V_{G} K\,;
\ee
here we have made use of the seesaw 
$M_\nu^M=-D_u U_\nu^\dagger D_\nu^{-1} U_\nu^* D_u$ written in the basis where the light neutrino mass matrix is diagonal, i.e., $m_{LL}=U_\nu^T D_\nu U_\nu$. This inequality can be interpreted as a necessary condition the $U_{\nu}$ matrix governing the formulae (\ref{gamma3})-(\ref{gamma5}) must obey in any perturbative and potentially realistic realization of the Witten's mechanism in the scheme of our interest.  

\subsection{CP conserving setting}
\subsubsection{The parameter space}

\begin{figure}[t]
\includegraphics[width=6cm]{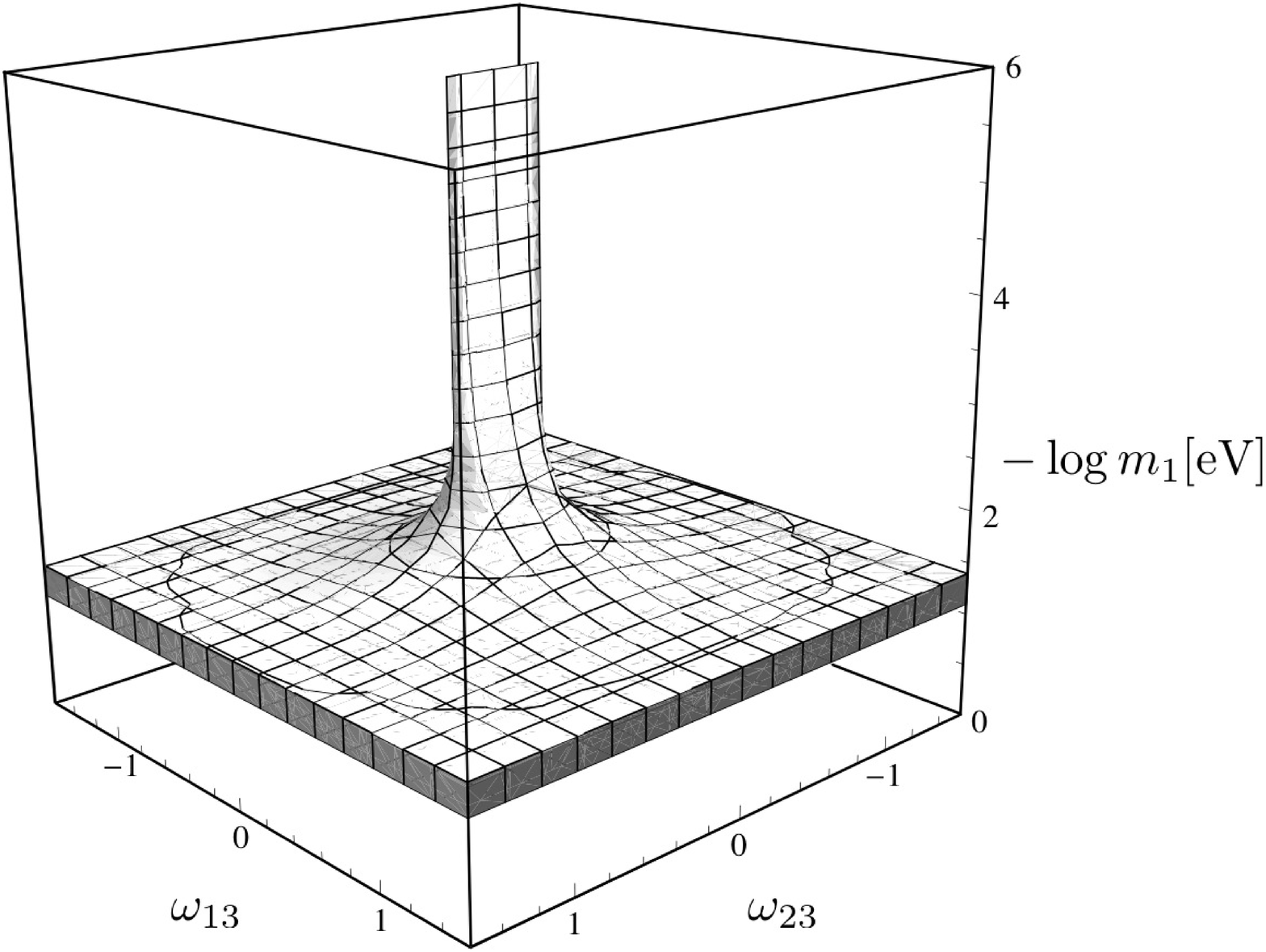}\hskip 5mm
\includegraphics[width=6cm]{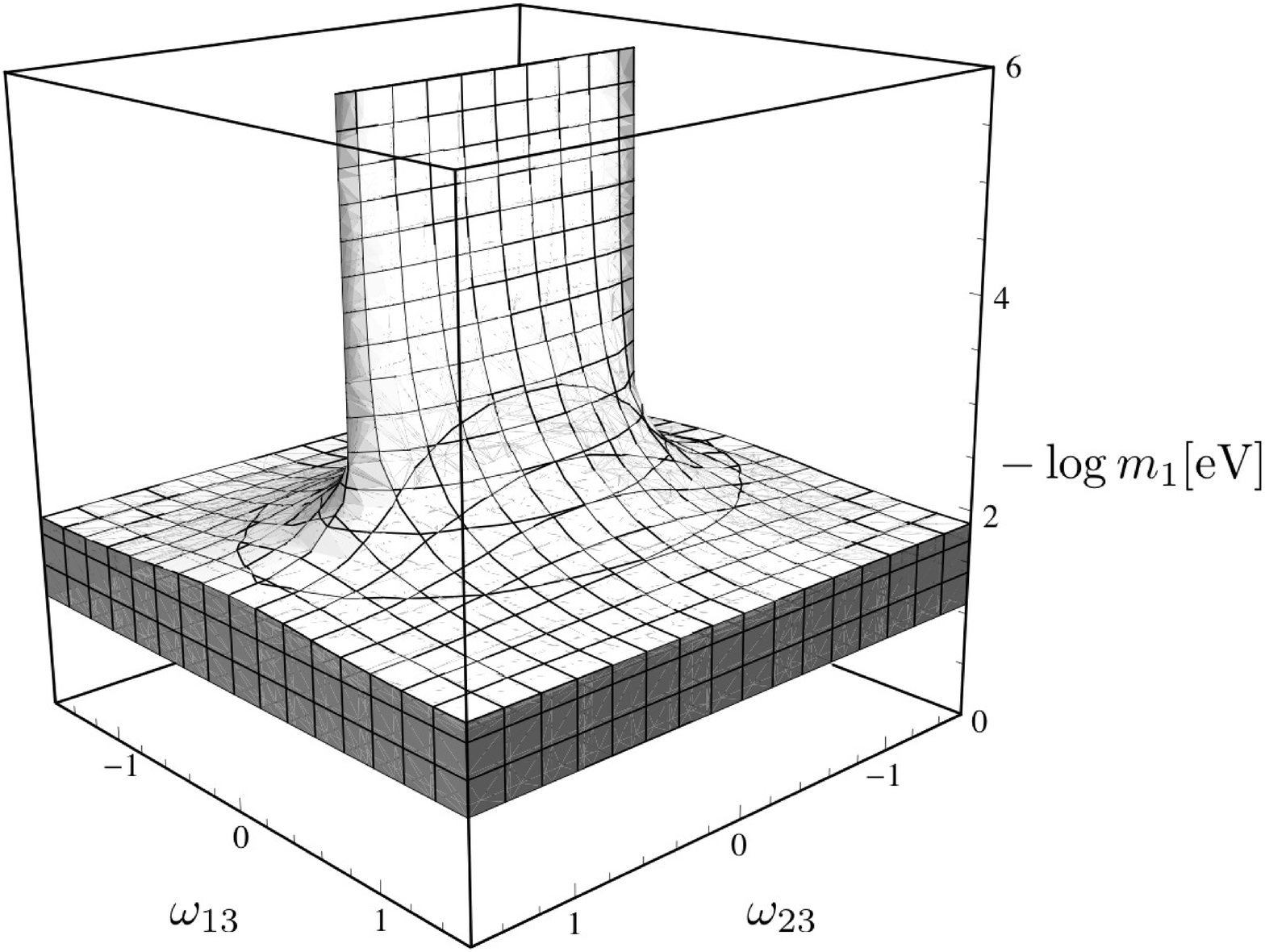}
\caption{\label{fig:domains}The shapes of the 
 $\omega_{23}$ - $\omega_{13}$ - $m_{1}$ space compatible with the SM vacuum stability and perturbativity constraints discussed in the text, for $K=10$ (left) and $K=30$ (right), respectively. The straight cuts in the lower part of the plot discarding the $m_{1}\gtrsim 8\times 10^{-2}$~eV regions corespond to the current cosmology limits~\cite{Riemer-Sorensen:2013jsa}. For small $m_{1}$ only a compact domain for  $\omega_{23}$ and $\omega_{13}$ is allowed; this, subsequently, gives rise to the constraints on the BNV observables.}
\end{figure}

For the sake of simplicity, let us start with the CP conserving setting, i.e., let us assume that the $U_{\nu}$ and $\PMNS$ matrices are real. 
It is easy to see that for non-negligible lightest neutrino mass $m_{1}$ (assuming normal hierarchy, cf.~\cite{Rodriguez:2013rma}) the LHS of Eq.~(\ref{centralformula}) is dominated by the 33 element. 
Using the ``standard CKM'' parametrization for $U_{\nu}$, i.e.,
$
U_\nu=U_{2\text{-}3}(\omega_{23})U_{1\text{-}3}(\omega_{13})U_{1\text{-}2}(\omega_{12})
$ where $U_{i\text{-}j}(\omega_{ij})$ stands for a rotation in the $i$-$j$ plane by an angle $\omega_{ij}$, e.g.
\be\label{Utheta}
U_{2\text{-}3}(\omega_{23})=\left(\begin{array}{ccc}
1&0&0\\
0&\cos{\omega_{23}}&\sin{\omega_{23}}\\
0&-\sin{\omega_{23}}&\cos{\omega_{23}}
\end{array}\right)\,,
\ee
and taking, for illustration, $V_G=10^{16}\,\mathrm{GeV}$ and $g_{G}=0.5$, i.e., $\alpha_G=0.02$,
Eq.~(\ref{centralformula}) can be rewritten as
\be\label{el33}
\frac{\sin^2\omega_{13}}{m_1} + \cos^2\omega_{13}\left(\!\frac{\sin^2\omega_{23}}{m_2}+\frac{\cos^2\omega_{23}}{m_3}\!\right)\leq K\times 3 \, \mathrm{eV}^{-1},
\ee
which, remarkably enough, is $\omega_{12}$-independent. Let us reiterate that for each $m_{1}$ this inequality defines an allowed domain in the $\omega_{23}$-$\omega_{13}$ space which conforms the SM stability and perturbativity constraints. Two examples of such allowed domains for $K=10$ and $K=30$ are depicted in FIG.~\ref{fig:domains}. It is important to notice that for small-enough $m_{1}$ the allowed region is compact, i.e., not all $\omega_{13}$ and, in particular, not all $\omega_{23}$ are allowed. This is the core of the argument which, subsequently, leads to distinctive features in the BNV observables of our interest. 

\subsubsection{Observables}
Barring isospin symmetry, there are two independent observables that may be exploited in the $d=6$ $p$-decay channels to neutral mesons, namely, $\Gamma(p\to \pi^{0}e^{+})\equiv \Gamma_{e}$ and $\Gamma(p\to \pi^{0}\mu^{+})\equiv \Gamma_{\mu}$. Needless to say, for each $U_{\nu}$ both these quantities are fully fixed; however, as we saw above, the SM vacuum stability and perturbativity constraints fix only $\omega_{13}$ and $\omega_{23}$ while leaving $\omega_{12}$ unconstrained. Hence, in order to get robust results, one should always consider extrema of the desired observables along the $\omega_{12}$ direction, i.e., optimize with respect to that angle. This is facilitated by the $\omega_{12}$-linearity of the relevant matrix elements; thus, such an optimization can be done analytically.  

\begin{figure}
\includegraphics[width=16cm]{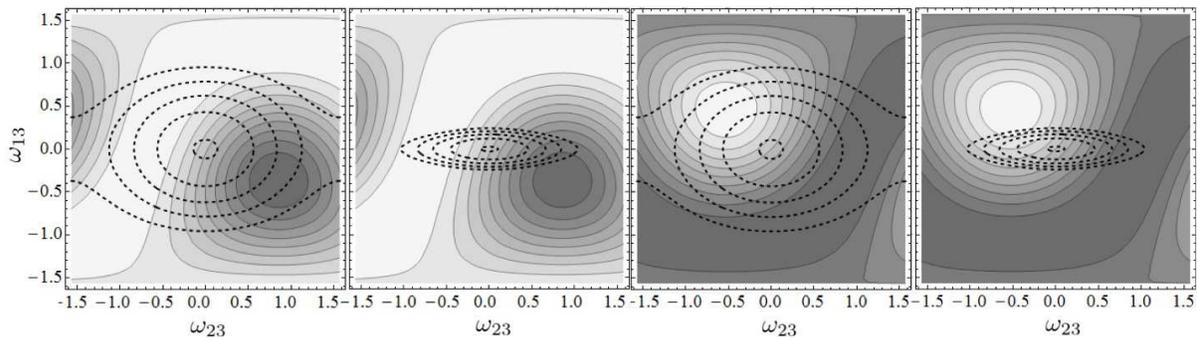}
\caption{\label{contours}Upper limits $\Gamma_{\mu}^{<}(\omega_{13},\omega_{23})$ (left two panels) and lower limits $(\Gamma_{e}+\Gamma_{\mu})^{>}(\omega_{13},\omega_{23})$ (right two panels) normalized to $\Gamma(p\to \pi^+\overline{\nu})|(V_{CKM})_{11}|^2/2$ as functions of $\omega_{13}$ and $\omega_{23}$ with values from 1 (lightest) to 0 (darkest color), see the text. In dashed lines we superimpose the boundaries of the domains consistent with the SM vacuum stability and parturbativity requirements for $K$'s from 7 (innermost contours) to $K=30$ (outermost contours) and for $m_{1}=0.8\times 10^{-2}$ eV (the first and third pannel) and $m_{1}=0.8\times 10^{-3}$ eV (the second and fourth pannel). The global extrema of the two observables of interest are displayed in FIG.~\ref{numerics}.}
\end{figure}

\begin{figure}
\includegraphics[width=14cm]{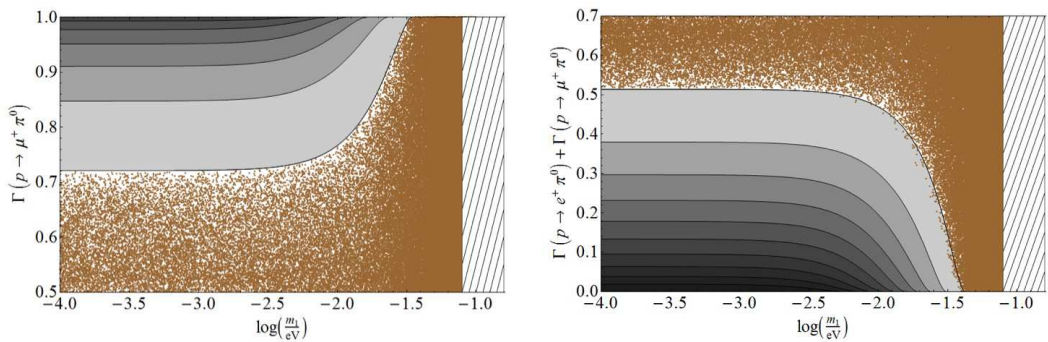}
\caption{\label{numerics}The global extrema of $\Gamma_{\mu}$ and $\Gamma_{e}+\Gamma_{\mu}$ as functions of $m_{1}$ for $K=7$ (the lowest line on the left panel and the highest line in the right panel, respectively) with each consecutive line corresponding to $K$ increased by 2. The results of a dedicated numerical analysis for $K=7$ have been superimposed over the relevant regions in brown points. The cuts on the right correspond to the global limit on the mass of the lightest neutrino derived from cosmology, see, e.g.,~\cite{Riemer-Sorensen:2013jsa} and references therein.}
\end{figure}

Starting with $\Gamma_{\mu}$, an upper limit $\Gamma_{\mu}^{\rm <}(\omega_{13},\omega_{23})$ (for any given $\omega_{13}$ and $\omega_{23}$) is attained for  $\tan{\omega_{12}^{\rm opt}}=\PMNSx_{23}\sin{\omega_{23}}-\PMNSx_{22}\cos{\omega_{23}}/\left[\PMNSx_{21}\cos{\omega_{13}}-\sin{\omega_{13}}\left(\PMNSx_{23}\cos{\omega_{23}}+\PMNSx_{22}\sin{\omega_{23}}\right)\right]$ where $V$ is a shorthand for $\PMNS$.
On the other hand, varying $\omega_{12}$, no such feature is observed in $\Gamma_{e}$ as it tends to cover the whole allowed region. However, there turns out to be a strong (anti-)correlation among the two and if one considers $\Gamma_{e}+\Gamma_{\mu}$ instead, a distinctive feature (a lower limit $(\Gamma_{e}+\Gamma_{\mu})^{>}(\omega_{13},\omega_{23})$) emerges again. In this case, the extremum along the $\omega_{12}$ direction is attained for $\tan{\omega_{12}^{\rm opt}}=\PMNSx_{33}\sin{\omega_{23}}-\PMNSx_{32}\cos{\omega_{23}}/\left[\PMNSx_{31}\cos{\omega_{13}}-\sin{\omega_{13}}\left(\PMNSx_{33}\cos{\omega_{23}}+\PMNSx_{32}\sin{\omega_{23}}\right)\right]$.
These limits, i.e., $\Gamma_{\mu}^{\rm <}(\omega_{13},\omega_{23})$ and $(\Gamma_{e}+\Gamma_{\mu})^{>}(\omega_{13},\omega_{23})$, are depicted in FIG.~\ref{contours} as functions of $\omega_{13}$ and $\omega_{23}$. Finally, if these $\omega_{13},\omega_{23}$-dependent quantities are superimposed with the consistency constraints discussed in the previous section, one obtains robust predictions for the quantities of interest. 
The global maxima of $\Gamma_{\mu}^{\rm <}(\omega_{13},\omega_{23})$ and the global minima of $(\Gamma_{e}+\Gamma_{\mu})^{>}(\omega_{13},\omega_{23})$ over the whole allowed parameter space for a given $m_{1}$ and a set of sample values of the $K$ parameter are depicted in FIG.~\ref{numerics}. For the sake of simplicity, $\omega_{13}$ has always been fixed at zero which, as one can see in FIG.~\ref{contours}, is a very good approximation, especially for smaller $m_{1}$. In the same plot, we display the results of a dedicated numerical analysis of the same problem without any extra assumption on $\omega_{13}$ which further confirms the validity of the simplified analytic approach.

\subsection{CP violating setting}
\begin{figure}[th]
\includegraphics[width=6cm]{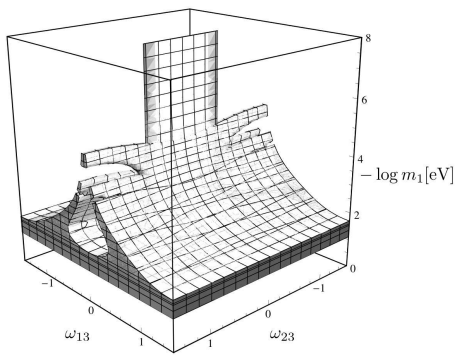}
\caption{\label{chimneyCPV}A typical shape of the parameter space allowed by the SM vacuum stability and perturbativity in  the CP-violating case, i.e., for  complex $U_{\nu}$ and $\PMNS$, cf. also FIG.~\ref{fig:domains}. In the displayed case the ``Dirac'' CP phase of $U_{\nu}$ was set to maximal, i.e., $\sigma=\pi/2$.}
\end{figure}

Even for complex $U_{\nu}$ (and $V_{PMNS}$), the logic of the argument remains the same. Some of the extra phases, i.e., those that may be factorized out of the product $\PMNS U_{\nu}$, are trivially harmless in formulae (\ref{gamma3})-(\ref{gamma5}) and we shall ignore them. Remarkably, this is also the case with the majority of the remaining phases therein with the only exception of the ``Dirac'' phase $\sigma$ in $U_{\nu}$ (i.e., the one corresponding to the $\delta$ CP phase in the CKM matrix); if large (i.e., close to $\pi/2$) $\sigma$  is allowed, the narrow ``chimney''-like shape of the allowed parameter space (cf. FIG.~\ref{fig:domains}) is disturbed until $m_{1}$ is rather small, see FIG.~\ref{chimneyCPV}. In such a case, the features in $\Gamma_{\mu}$ and $\Gamma_{e}+\Gamma_{\mu}$ tend to be smeared. For a more detailed discussion of the CP violating case with numerical illustrations the reader is deferred to the more detailed study~\cite{Rodriguez:2013rma}. 

\section{The minimal potentially realistic model}
Attentive readers have certainly noticed that, so far, we have left aside an extra piece of information the flavour structure of the minimal flipped $SU(5)$ supplies, namely, the correlation between $M_{\nu}^{M}$ and $M_{d}$ which are both proportional to the (symmetric) Yukawa matrix $Y_{10}$. This is slightly unfortunate because the light neutrino spectrum in such a case turns out to be too hierarchical, typically $m_{2}:m_{3}\sim 1\permil$ rather than the desired ${\cal O}(10\%)$. However, there is a trivial way to break this correlation while preserving all the desired features of the simplest setting (namely, $M_{d}=M_{d}^{T}$ without which the $d=6$ proton decay may be ``rotated out'' to a large degree~\cite{Nath:2006ut,Dorsner:2004xx,Barr:2013gca} and also $M_{\nu}^{D}=M_{u}^{T}$ that was crucial for the derivation of the key formula (\ref{centralformula})). It consists in adding an extra copy of $5_{H}$ (to be denoted $5_{H}'$) to the scalar sector of the theory. The generalized Yukawa Lagrangian 
 \bea\label{welcome3}
{\cal L}_{Y}&\ni &Y_{10}10_{M}10_{M}5_{H}+Y'_{10}10_{M}10_{M}5'_{H}+Y_{\overline{5}}10_{M}\overline{5}_{M}{5}^{*}_{H}+Y'_{\overline{5}}10_{M}\overline{5}_{M}{5'_{H}}^{*}+Y_{1}\overline{5}_{M}1_{M}5_{H}+Y'_{1}\overline{5}_{M}1_{M}5_{H}'
+h.c.
\eea
then yields the following set of sum-rules for the effective quark and lepton mass matrices:
\be
M_{\nu}^{D}=M_{u}^{T}\propto Y_{\overline{5}}v_{5}^{*}+Y'_{\overline{5}}v_{5'}^{*},  \quad
M_{d}= M_{d}^{T}=Y_{10}v_{5}+Y'_{10}v_{5'}\label{Mdmain}
\ee
\be
M_{e}= Y_{1}v_{5}+Y'_{1}v_{5'} \quad \text{arbitrary.}
\ee
At the same time, there is a second Yukawa matrix  entering linearly the Witten's formula (\ref{MnuMmain}) which may be thus rewritten as
\be
M_{\nu}^{M}= \left(\frac{1}{16\pi^{2}}\right)^{2}g^{4}(Y_{10}\,\mu+Y_{10}'\,\mu')\frac{\vev{10_{H}}^{2}}{M_{G}^{2}}\times {\cal O}(1)\label{MnuMmainGeneralized}\,.
\ee
Hence, the unwelcome correlation between $M_{\nu}^{M}$ and $M_{d}$ is alleviated. Moreover, the only technical difference between (\ref{MnuMmainGeneralized}) and (\ref{MnuMmain}) is the presence of an extra ``$\mu Y_{10}$'' factor on the RHS of Eq.~(\ref{MnuMmainGeneralized}) which, barring accidental cancellations, can be accounted for by a mere doubling of the RHS of Eq.~(\ref{centralformula}). Hence, it is very easy to adopt all the results obtained in the previous section for the simplest model to the fully realistic case with a pair of scalar $5_{H}$'s; for example, the allowed points in FIG.~\ref{numerics} for $K=8$ are allowed in the generalized setting with $K=4$ and so on.
\section{Conclusions and outlook}
We have argued that the Witten's mechanism for the radiative RH neutrino mass generation originally identified in the realm of the simplest $SO(10)$ grand unifications can be easily adopted to the flipped $SU(5)$ framework. In such a case, it strongly benefits from the relaxed gauge coupling unification constraints on its key ingredient, namely, the rank-breaking VEV of the relevant scalar field which, unlike in the non-SUSY $SO(10)$ GUTs, tends to be as high as $10^{16}$ GeV. This, due to the inherent double loop suppression leads to the RH neutrino mass scale in the $10^{13}$ GeV ballpark which, in seesaw, is just right for the light neutrino masses in the sub-eV domain. Moreover, the tight correlation between the lepton and quark sectors inherent to essentially all unifications leads to distinctive BNV signals which may be within reach of the future megaton-scale proton-decay/neutrino facilities such as LBNE and/or Hyper-Kamiokande.

\subsection{Acknowledgments}
The work of M.M. is supported by the Marie-Curie Career Integration Grant within the 7th European Community Framework Programme
FP7-PEOPLE-2011-CIG, contract number PCIG10-GA-2011-303565 and by the Research proposal MSM0021620859 of the Ministry of Education, Youth and Sports of the Czech Republic. The work of H.K. is supported by the Grant Agency of the Czech Technical University in Prague, grant No. SGS13/217/OHK4/3T/14. The work of C.A.R. is in part supported by EU Network grant UNILHC PITN-GA-2009-237920 and by the Spanish MICINN grants FPA2011-22975, MULTIDARK CSD2009-00064 and the Generalitat Valenciana (Prometeo/2009/091). M.M. is indebted to the organisers of the marvellous CETUP'13 for hospitality and support.


\begin{thebibliography}{10}

\bibitem{Minkowski:1977sc}
P.~Minkowski,
\newblock Phys. Lett. {\bf B67}, 421 (1977).

\bibitem{Yanagida:1979as}
T.~Yanagida,
\newblock Horizontal gauge symmetry and masses of neutrinos,
\newblock in {\em Proc. Workshop on the Baryon Number of the Universe and
  Unified Theories}, edited by O.~Sawada and A.~Sugamoto, p.~95, 1979.

\bibitem{Mohapatra:1979ia}
R.~N. Mohapatra and G.~Senjanovic,
\newblock Phys. Rev. Lett. {\bf 44}, 912 (1980).

\bibitem{Schechter:1980gr}
J.~Schechter and J.~W.~F. Valle,
\newblock Phys. Rev. {\bf D22}, 2227 (1980).

\bibitem{Lazarides:1980nt}
G.~Lazarides, Q.~Shafi, and C.~Wetterich,
\newblock Nucl. Phys. {\bf B181}, 287 (1981).

\bibitem{Foot:1988aq}
R.~Foot, H.~Lew, X.~He, and G.~C. Joshi,
\newblock Z.Phys. {\bf C44}, 441 (1989).

\bibitem{Zee:1980ai}
A.~Zee,
\newblock Phys.Lett. {\bf B93}, 389 (1980).

\bibitem{Zee:1985rj}
A.~Zee,
\newblock Phys.Lett. {\bf B161}, 141 (1985).

\bibitem{Zee:1985id}
A.~Zee,
\newblock Nucl.Phys. {\bf B264}, 99 (1986).

\bibitem{Babu:1988ki}
K.~Babu,
\newblock Phys.Lett. {\bf B203}, 132 (1988).

\bibitem{Bonnet:2012kz}
F.~Bonnet, M.~Hirsch, T.~Ota, and W.~Winter,
\newblock JHEP {\bf 1207}, 153 (2012), arXiv:1204.5862 [hep-ph].

\bibitem{Angel:2012ug}
P.~W. Angel, N.~L. Rodd, and R.~R. Volkas,
\newblock Phys.Rev. {\bf D87}, 073007 (2013), arXiv:1212.6111 [hep-ph].

\bibitem{Babu:2013pma}
K.~Babu and J.~Julio,
\newblock arXiv:1310.0303 [hep-ph].

\bibitem{Baek:2012ub}
S.~Baek, P.~Ko, and E.~Senaha,
\newblock (2012), arXiv:1209.1685 [hep-ph].

\bibitem{Ohlsson:2009vk}
T.~Ohlsson, T.~Schwetz, and H.~Zhang,
\newblock Phys.Lett. {\bf B681}, 269 (2009), arXiv:0909.0455 [hep-ph].

\bibitem{Nebot:2007bc}
M.~Nebot, J.~F. Oliver, D.~Palao, and A.~Santamaria,
\newblock Phys.Rev. {\bf D77}, 093013 (2008), arXiv:0711.0483.

\bibitem{AristizabalSierra:2006gb}
D.~Aristizabal~Sierra and M.~Hirsch,
\newblock JHEP {\bf 0612}, 052 (2006), arXiv:hep-ph/0609307.

\bibitem{Frampton:2001eu}
P.~H. Frampton, M.~C. Oh, and T.~Yoshikawa,
\newblock Phys.Rev. {\bf D65}, 073014 (2002), arXiv:hep-ph/0110300.

\bibitem{Abe:2011ts}
K.~Abe {\em et~al.},
\newblock (2011), arXiv:1109.3262 [hep-ex].

\bibitem{Akiri:2011dv}
LBNE Collaboration, T.~Akiri {\em et~al.},
\newblock (2011), arXiv:1110.6249 [hep-ex].

\bibitem{Autiero:2007zj}
D.~Autiero {\em et~al.},
\newblock JCAP {\bf 0711}, 011 (2007), arXiv:0705.0116 [hep-ph].

\bibitem{Witten:1979nr}
E.~Witten,
\newblock Phys. Lett. {\bf B91}, 81 (1980).

\bibitem{Chang:1984qr}
D.~Chang, R.~N. Mohapatra, J.~Gipson, R.~E. Marshak, and M.~K. Parida,
\newblock Phys. Rev. {\bf D31}, 1718 (1985).

\bibitem{Deshpande:1992au}
N.~G. Deshpande, E.~Keith, and P.~B. Pal,
\newblock Phys. Rev. {\bf D46}, 2261 (1993).

\bibitem{Deshpande:1992em}
N.~G. Deshpande, E.~Keith, and P.~B. Pal,
\newblock Phys. Rev. {\bf D47}, 2892 (1993), arXiv:hep-ph/9211232.

\bibitem{Bertolini:2009qj}
S.~Bertolini, L.~Di~Luzio, and M.~Malinsky,
\newblock Phys. Rev. {\bf D80}, 015013 (2009), arXiv:0903.4049 [hep-ph].

\bibitem{Bajc:2004hr}
B.~Bajc and G.~Senjanovic,
\newblock Phys. Lett. {\bf B610}, 80 (2005), hep-ph/0411193.

\bibitem{Bajc:2005aq}
B.~Bajc and G.~Senjanovic,
\newblock Phys.Rev.Lett. {\bf 95}, 261804 (2005), arXiv:hep-ph/0507169.

\bibitem{Derendinger:1983aj}
J.~Derendinger, J.~E. Kim, and D.~V. Nanopoulos,
\newblock Phys.Lett. {\bf B139}, 170 (1984).

\bibitem{PhysRevLett.45.413}
A.~De~R\'ujula, H.~Georgi, and S.~L. Glashow,
\newblock Phys. Rev. Lett. {\bf 45}, 413 (1980).

\bibitem{Barr:1981qv}
S.~M. Barr,
\newblock Phys. Lett. {\bf B112}, 219 (1982).

\bibitem{Rodriguez:2013rma}
C.~A. Rodr{\'\i}guez, H.~Kole{\v s}ov{\'a}, and M.~Malinsk{\'y},
\newblock arXiv:1309.6743 [hep-ph].

\bibitem{Nath:2006ut}
P.~Nath and P.~F. Perez,
\newblock Phys. Rept. {\bf 441}, 191 (2007), hep-ph/0601023.

\bibitem{Das:2005eb}
C.~Das, C.~Froggatt, L.~Laperashvili, and H.~Nielsen,
\newblock Mod.Phys.Lett. {\bf A21}, 1151 (2006), arXiv:hep-ph/0507182.

\bibitem{Abel:1989hq}
S.~Abel,
\newblock Phys.Lett. {\bf B234}, 113 (1990).

\bibitem{Babuprivateconversation}
Private discussion with K.S. Babu.

\bibitem{Buttazzo:2013uya}
D.~Buttazzo {\em et~al.},
\newblock arXiv:1307.3536 [hep-ph].

\bibitem{Riemer-Sorensen:2013jsa}
S.~Riemer-S{\o}rensen, D.~Parkinson, and T.~M. Davis,
\newblock arXiv:1306.4153 [astro-ph].

\bibitem{Dorsner:2004xx}
I.~Dorsner and P.~Fileviez~Perez,
\newblock Phys.Lett. {\bf B605}, 391 (2005), arXiv:hep-ph/0409095.

\bibitem{Barr:2013gca}
S.~Barr,
\newblock arXiv:1307.5770 [hep-ph].

\end{thebibliography}
\end{document}